\documentclass{JHEP3}
\usepackage{epsfig}
\usepackage{amssymb}

\newcommand{\bx}{{\mathbf x}}

\newcommand{\al}{\alpha}

\newcommand{\de}{\delta}
\newcommand{\De}{\Delta}

\newcommand{\ga}{\gamma}

\newcommand{\la}{\lambda}

\newcommand{\om}{\omega}

\newcommand{\ra}{\rightarrow}

\newcommand{\be}{\begin{equation}}
\newcommand{\ee}{\end{equation}}
\newcommand{\bea}{\begin{eqnarray}}
\newcommand{\eea}{\end{eqnarray}}
\newcommand{\bean}{\begin{eqnarray*}}
\newcommand{\eean}{\end{eqnarray*}}
\newcommand{\dd}{\partial}

\title{
Superluminal motion and closed signal curves
}
\author{Camille Bonvin \\ D\'epartement
de Physique Th\'eorique, Universit\'e de
  Gen\`eve, 24 quai Ernest Ansermet, CH--1211 Gen\`eve 4, Switzerland\\ E-mail:
\email{camille.bonvin@physics.unige.ch}}

\author{Chiara Caprini \\ D\'epartement de Physique Th\'eorique, Universit\'e
de Gen\`eve, 24 quai Ernest Ansermet, CH--1211 Gen\`eve 4,
Switzerland \\ E-mail:\email{chiara.caprini@physics.unige.ch}}

\author{Ruth Durrer \\ D\'epartement de
Physique Th\'eorique, Universit\'e de
Gen\`eve, 24 quai Ernest Ansermet, CH--1211 Gen\`eve 4, Switzerland\\ 
E-mail:\email{ruth.durrer@physics.unige.ch}}

\abstract{
We discuss some properties of Lorentz invariant theories which allow
for superluminal motion. We show that, if signals are always sent
forward in time, closed curves along which signals propagate can
be formed. This leads to problems with causality and with the second
law of thermodynamics. Only if one singles out one frame with
respect to which all signals travel forward in time, the formation
of 'closed signal curves' can be prevented. However, the price to pay is 
that in some reference frames perturbations propagate towards the past or
towards the future, depending on the direction of emission.}


\preprint

\begin{document}

\section{Introduction}
\label{sec:intro}

It is a matter of debate in the literature whether a theory that
admits superluminal propagation is acceptable
\cite{Hashimoto,Liberati,Gib,AA,Bruneton,Dubovsky,Babichev1,Ellisetal,Babichev2}.
It has been argued that superluminal motion needs not lead to
closed 'timelike' curves, and is therefore not problematic.
Furthermore, it has been put forward that perturbations on a
background which is not Lorentz invariant (i.e., around a solution
of the equations of motion which breaks Lorentz invariance) can
very well propagate faster than the speed of light, without
leading to serious problems with causality.

In this paper we first show that whenever the Lagrangian
for a field is such that field modes can propagate at superluminal 
speeds, closed curves along which a signal propagates can be
constructed. We call them 'closed signal curves' or short CSC's.
In a next step we show that, for a fixed cosmological background
solution, the same result holds if one requires that observers can
send signals only forward in time, i.e., a forward time direction
exists unambiguously in each reference frame. Only if we require
that all signals, independently from the frame with respect to
which they have been emitted, travel forward with respect to the
time  of the cosmological  reference frame, we can avoid the
possibility of CSC's. However, this goes at the cost that
observers traveling at high (but sub-luminal) speed with respect
to the cosmological frame must send signals backwards in their
time for some specific directions. In other words, fluctuations in
these frames propagate sometimes with the advanced and sometimes
with the retarded Green function.

It seems clear to us, that in a universe with closed signal
curves, physics, as we know it, is no longer possible. 
For example, the second law of thermodynamics is violated, 
since after one turn in a closed loop, the original state of the system
must be re-established hence entropy cannot have grown. If these
loops are of Planckian size or much larger than the age of the
universe, there may be a way out of contradiction with every day
experience, but if the loops can be of mm or cm size, this becomes
very difficult. Especially, it is not clear to us whether
thermodynamics in general, and the concept of entropy in particular,
can still make sense in a universe with closed signal curves.

The point of the present note is to show that theories which do admit
superluminal motion, either admit closed signal curves or
force some observers to send signals backwards in their time.
This finding is independent of the fact whether or not the
background breaks Lorentz invariance.

In the next section we construct closed signal curves in a field
theory which allows for superluminal motion. We discuss our result
and show that it can be avoided by additional assumptions if
we have a preferred reference frame, like in cosmology. We also
formulate the conditions under which scalar field Lagrangians allow
superluminal motion. In Section~III we discuss in more detail the
cosmological situation concentrating especially on the example of
k-essence~\cite{ke1,ke+,picon,kess} and in Section~IV we conclude. The
speed of light is $c=1$ and we use the metric signature
$(+,-,-,-)$.

\section{Closed signal curves from superluminal velocities}
\label{sec:closed}

It is well known that covariant Lagrangians can lead to
superluminal motion. To be specific and to simplify matters, let
us consider the Lagrangian of a scalar field $\phi$, leading to a
covariant equation of motion of the form
\be\label{eqmo}
G^{\mu\nu}\nabla_\mu\nabla_\nu\phi = {\rm lower~order~terms}~,
\ee
where $G^{\mu\nu}$ is a symmetric
tensor field given by $\phi$ and other degrees of freedom. It need
not be the spacetime metric. If $G^{\mu\nu}$ is non-degenerate and has
Lorentzian signature, Eq.~(\ref{eqmo}) is a hyperbolic equation of
motion. We assume this to be the case (see \cite{picon} for a discussion about this issue). 
The null-cone of the co-metric $G^{\mu\nu}$ is the characteristic cone of this
equation. The rays are defined by the 'metric' $(G^{-1})_{\mu\nu}$
such that $G^{\mu\nu}(G^{-1})_{\nu\la}={\de^\mu}_\la$. The
characteristic cone limits the propagation of field modes in the
sense that the value of the field at some event $q$ is not
affected by the values outside the past characteristic cone and,
on the other hand, that the value at $q$ cannot influence the
field outside the future characteristic cone~\cite{CH}.

For very high frequencies, the lower order terms are subdominant and
the field propagates along the characteristic cone. At lower
frequencies, lower order terms act similarly to an effective mass
and the field propagates inside the characteristic cone. We now show that,
closed signal curves can be constructed if this cone is wider than
the light cone defined by the spacetime metric $g_{\mu\nu}$.

If the characteristic cone of  $G_{\mu\nu}$ is wider than the light
cone, the maximal propagation velocity $v_{\max}$ of the field $\phi$, which
satisfies $G_{\mu\nu}v_{\max}^\mu v_{\max}^\nu =0$, is spacelike
(with respect to $g_{\mu\nu}$). Since
the notion 'spacelike' is frame independent, this is true in every
reference frame. Of course, the characteristic cone for $\phi$
is not invariant
under Lorentz transformations, but the fact that it is spacelike is.

We consider two reference frames $R$ and $R'$ with common origin $q_0$:
$(0,0)= (t,x)=(t',x')$. $R'$ is boosted with
respect to $R$ in $x$-direction with velocity $v<1$.
For an event $q$ with coordinates $(t,x)$ in $R$ and $(t',x')$ in $R'$
we have the usual transformation laws
\bea
t' &= \ga(t-vx)\, , \quad x' &= \ga(x-vt), ~ \ga
=\frac{1}{\sqrt{1-v^2}}~, \nonumber \\
t &= \ga(t'+vx')\, , ~ x &= \ga(x'+vt')~.
\eea
We assume that $v$ is sufficiently large such that the superluminal
velocity $1/v$ in
$x$-direction is inside the characteristic cone of $\phi$. We can then send
a $\phi$-signal from $(0,0)$ with speed $v_1>1/v$ in $x$-direction to the event
$q_1$, see Fig.~\ref{fig1}. The signal is received in $x_1$ at
time $t_1=x_1/v_1$. In $R'$ this event has the
coordinates $t_1'=
\ga t_1(1-vv_1)<0$ and $x_1'=\ga t_1 (v_1-v) >0$. Note also that
$v_1'=x_1'/t_1'=(v_1-v)/(1-vv_1)<0$, and the signal is propagating
into the past of $R'$.

We can choose $x_1$ and correspondingly $t_1$ very small so that
curvature is negligible on these scales and we may identify the
spacetime manifold with its tangent space at $(0,0)$. In other words
we want to choose these dimensions sufficiently small so that we may
neglect the position dependence of both, the light cone and the
characteristic cone for $\phi$. The situation is
then exactly analogous to the one of special relativity.

An observer in the frame $R'$ now
receives the signal emitted at $(0,0)$ in $R$ and returns it with
velocity $v_2'$ to $x_2=0=\ga(x_2'+vt_2')$. We denote the arrival
event by $q_2$. It has the coordinates $(t_2,0)$ with respect to $R$
and $(t_2',x_2')$ with respect to $R'$. Since $v_2' \neq v_1'$
in order for a CSC to form, we may have to transform the signal to
another frequency to allow it to travel with speed $v_2'$. If the
returned signal arrives at some time $t_2<0$, the observer
in $R$ which has received the signal simply stores it until the time
$|t_2|$ has elapsed after which this signal has propagated along the closed
curve $q_0 \ra q_1 \ra q_2 \ra q_0$ shown in Fig.~\ref{fig1}, a CSC
has been generated. 

Let us elaborate on the requirement $t_2<0$. We assume here that an
arbitrary observer can send signals only into her future, so that
$t_2'>t_1'$. Hence we want to choose $v_2'$ such that even though $\De
t' = t_2'-t_1'>0$, we have $\De t = t_2-t_1<0$. When sending a
signal with speed $v_2$ in frame $R$, respectively $v_2'$ in $R'$,
the times which elapse while the signal travels a distance $\De x$
respectively $\De x'$ are related by ($\De x'=v_2'\De t'$)
\bea
\De t &=& t_2-t_1 =\ga(t_2'+vx'_2-t_1'-vx_1') = \ga(\De t'+v\De x')=
\ga(1+vv_2')\De t'~,\\
\De t' &=& \ga(1-vv_2)\De t ~.
\eea
In order to achieve $\De t<0$ and at the same time
$\De t'>0$ we need $vv_2>1$, hence $v_2>1/v>1$.

From Fig.~\ref{fig1} it is evident that $v_2$ which is the inverse of the slope
of the line from $q_2$ to $q_1$ is smaller than $v_1$ which
is the inverse of the slope from $q_0$ to $q_1$. This is also
obtained from
$$
0<v_2 = \frac{-x_1}{t_2-t_1}= \frac{x_1}{t_1-t_2} < \frac{x_1}{t_1}=v_1~.
$$
For the inequality sign we have used $t_2<0$.
Therefore, also $v_2$ is inside the characteristic cone
of $\phi$ and is admitted as a propagation velocity. Note
that since $v_2>1$ the distance between the events $q_1$ and $q_2$
is spacelike. Also in the reference frame $R'$,  $v_1'>v_2'$, but
both these velocities are negative hence for the
absolute values we have $|v_2'|>|v_1'|$.

\FIGURE[!t]{\epsfig{file=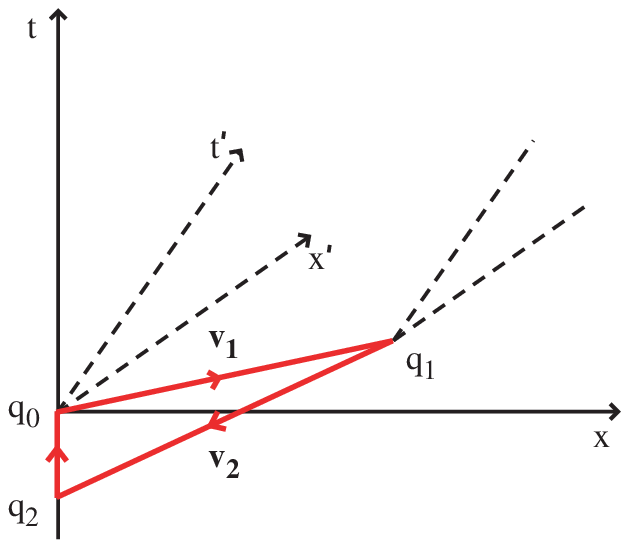,height=5.7cm}
\caption{\label{fig1} A closed signal curve going along $q_0\ra q_1\ra
  q_2\ra q_0$.}}

\subsection{Lagrangians which allow for superluminal motion}

We now identify scalar field Lagrangians which allow for superluminal
motion leading to the causal problem discussed above.
Consider a Lagrangian characterized by a non-standard
kinetic term, with the action
\be\label{action}
 S = \int d^4x \sqrt{-g}\left[- \frac{R}{6} + K(\phi)p(X) -V(\phi)\right]~,
\ee
were $\phi$ is the scalar field (for example, a tachyon \cite{Sen}, the
k-essence field \cite{ke1}, or the k-inflaton \cite{ke2}). 
$X=\frac{1}{2}\nabla_\mu\phi\nabla^\mu\phi$
is the kinetic energy; we use units with $\frac{8\pi G}{3}=1$.
The equation of motion for $\phi$ is given by
\be
Kp_{,X}G^{\mu\nu}\nabla_\mu\nabla_\nu\phi = -K_{,\phi}p
- V_{,\phi} - 2Xp_{,X}K_{,\phi} ~. 
\ee
The potential term and first order derivatives are irrelevant
for the characteristics of the field equation. These are given by the co-metric
$G^{\mu\nu}$. If  a prime denotes derivative with respect to $X$, the
co-metric is
\be
G^{\mu\nu}=g^{\mu\nu}+\frac{p''(X)}{p'(X)}\nabla^\mu\phi\nabla^\nu\phi~.
\ee
As discussed above,
for the signal not to propagate faster than the speed of light, the characteristic
cone should not lie outside the metric cone. This means that the unit normal to
the characteristics $S_\mu$ must not be timelike with respect to $g^{\mu\nu}$
\cite{Gib,bek}. The condition
\be
G^{\mu\nu}S_\mu S_\nu=0
\ee
implies
\be
g^{\mu\nu}S_\mu S_\nu=-\frac{p''(X)}{p'(X)}(\nabla^\mu\phi S_\mu)^2~.
\ee
Therefore, $S_\mu$ is not timelike if and only if
\be
\frac{p''(X)}{p'(X)}\geq 0~.
\ee
Every theory that does not fulfill this conditions runs into the
problem discussed above. Already in the 60ties,
the appearance of superluminal motion has led to the exclusion of
generic covariant higher spin $s\geq 1$
Lagrangians~\cite{VZ1}. Examples are the Lagrangian 
of a self-interacting neutral vector field, a minimally coupled spin $2$ field,
or the minimally coupled Rarita-Schwinger equation for a spin $3/2$ particle~\cite{VZ2}.

\section{Closed signal curves on a background}

So far, we have not specified any background upon which the $\phi$-signal
propagates. As we have seen above, the Lagrangian can be such that
the presence of a non-vanishing signal is sufficient for the
characteristic cone of $\phi$ to be spacelike or, equivalently,
its normal to be timelike, and hence the propagation to be superluminal.

On the other hand, one can consider the propagation of fluctuations
upon a fixed background $\phi_0$. If $\nabla\phi_0\neq 0$ is timelike, this
generates a preferred frame of reference, the one in which
$\nabla\phi_0$ is parallel to $t$. Let us call this reference
frame $R_0$. If the null-cone of the metric $G^{\mu\nu}(\phi_0)$
is spacelike (always with respect to the spacetime metric), the
construction  leading to a CSC presented in the previous section is
still possible. However, now there is in 
principle a way out. If we require that signals always propagate forward
in time in the frame $R_0$, closed signal
curves become impossible. The CSC $q_0 \ra q_1\ra q_2 \ra q_0$ is
also closed in $R_0$. As it encloses a non-vanishing area it must
contain both, a part where it advances in time and a part where it
goes backward in time, so that it violates the requirement that the
signal can only advance in time in the frame $R_0$.

This is the main point. In relativity, events with spacelike
separations have no well defined chronology. Depending on the
reference frame we are using, $q_2$ is either before (in $R$) or after
(in $R'$) $q_1$. If we can send a signal from $q_1$ to $q_2$, this
signal travels forward in time in $R'$ and backward in time in $R$. In
the frame which is boosted with respect to $R$ with velocity $1/v_2$ in
$x$-direction, the signal has even infinite velocity: $q_1$ and
$q_2$ have the same time coordinate in this reference frame.

If we require every signal to travel forward with respect to the time
coordinate of $R_0$, we shall no longer have closed curves along which
a signal propagates, but we then have signals propagating into
the past in the boosted reference frame $R'$ in which they have been emitted.
Moreover, the field value at some point $q$ can now
be influenced by field  values in the future. This may sound very
bizarre; however, as far as we can see, it is not contradictory since the
events in the future which can influence $q$ are in its spacelike
future and cannot be influenced by $q$. On the other hand, the events
in its past which $q$ can influence are in its spacelike past and
they cannot influence $q$, see Fig.~\ref{fig2}.
In the limit in which the maximal propagation velocity
$v_{\max}$ derived from Eq.~(\ref{eqmo}) approaches infinity in the
frame $R_0$, the past and future cones
in the boosted reference frame will approach each other, but never overlap.
The cone edge $x'=v_{\max}'t'$ is always flatter than the one $x'=(-v_{\max})'t'$: one has
$ (-v_{\max})' = -(v_{\max}+v)/(1+vv_{\max})$, and $v_{\max}'
= (v_{\max}-v)/(1-vv_{\max})$, which both tend to $-1/v$ in the limit $v_{\max}\ra \infty$.
The opening angle $\al$
between $v_{\max}'$ and $(-v_{\max})'$ is given by
\be
\al = \frac{2v_{\max}}{v_{\max}^2-1}\left(\frac{1-v^2}{1+v^2}\right) ~.
\ee
Hence $\al \rightarrow 0$ if either $v\ra 1$ or $v_{\max}\ra\infty$.

\FIGURE{\epsfig{file=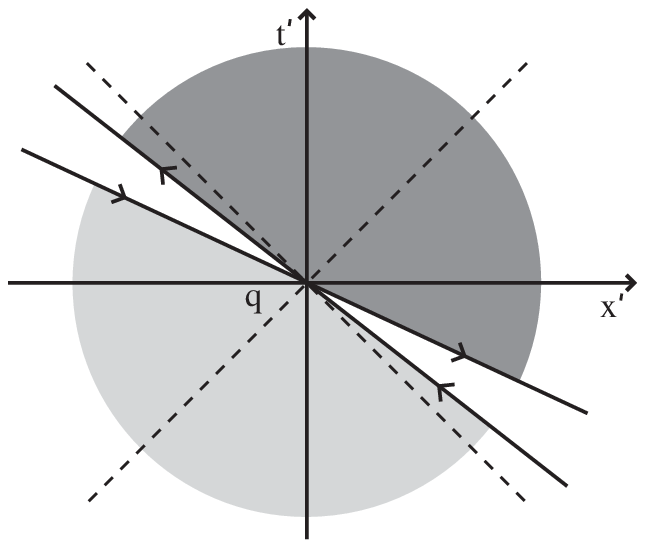,height=5.5cm}
\caption{\label{fig2} The characteristic cone $v_{\max}'$ and $(-v_{\max})'$ is shown with
arrows indicating the forward time direction in the preferred frame
$R_0$. All events inside the 'backward' characteristic cone
(light grey) can influence the event $q$, whereas $q$ can influence all
events inside the 'forward' characteristic cone (dark grey).}}

From Fig.~\ref{fig2} it is clear that there is no immediate contradiction
since there are no points which are simultaneously in the
past and future characteristic cone of $q$, hence no closed signal
curves or CSC's are possible. The physical interpretation is however quite
striking for an observer sitting at the origin of $R'$.
When sending a signal with a velocity close to $v_{\max}$ to
the left, it naturally propagates into the observer's future; when sending it
to the right, it has to propagate into her past, from where it can
reach her again later at a $t'>0$, when the past
cone from $(t',0)$ intersects the future cone
from $(0,0)$. However, also in the boosted frame $R'$, $\phi_0$ is
a solution of the equation of motion and were it not for the cosmological
symmetry, there would be no reason to prefer one frame over the other.

\subsection{k-essence}
We now consider in somewhat more detail the particular example of
k-essence~\cite{ke1,ke+}. We show that k-essence signals with different
wave numbers can propagate with different superluminal velocities.

The k-essence action is given by
\be\label{actionk}
 S = \int d^4x \sqrt{-g}\left[- \frac{R}{6} + \frac{p(X)}{\phi^2}\right]~,
\ee
where now $\phi$ is the k-essence field and again 
$X=\frac{1}{2}\nabla_\mu\phi\nabla^\mu\phi$.

In \cite{kess} it has been shown that, in every k-essence model which
solves the coincidence problem
and leads to a period of acceleration, the field has to propagate
superluminally during some stage of its evolution. Therefore, k-essence
can lead to the formation of CSC's.
As discussed above, CSC's can be constructed using two different
superluminal propagation velocities (see Fig.~\ref{fig1}). In particular,
in the frame $R_0$ where the background is homogeneous and isotropic we need
$v_2<v_1$. This can be achieved because the equation of motion of k-essence
perturbations contains an effective mass term which leads to dispersion.
Therefore, different wave-numbers propagate with different velocities.
In the following, we calculate the group velocity of the k-essence
perturbations using the WKB approximation.

We split the k-essence field in the cosmic background solution and a perturbation,
$\phi=\phi_0(t)+\de\phi(t,\bx)$. In longitudinal gauge the metric is
\be
ds^2=\Big(1+2\Psi(t,\bx)\Big)dt^2-a(t)^2\Big(1-2\Psi(t,\bx)\Big)\delta_{ij}dx^idx^j~,
\ee
where $a(t)$ is the scale factor and $\Psi(t,\bx)$ is the Bardeen potential.
We restrict our calculations to the case in which k-essence is
subdominant with respect to matter and radiation. This is the case when
k-essence evolves from the radiation fixed point to the de Sitter fixed point
(see for example Fig.~1 in~\cite{kess}). As it is shown in~\cite{kess}, during
this stage the sound velocity $c_s^2$ has to be larger than one.
The equation of motion of k-essence perturbations depends on the
choice of initial conditions. One possibility is to consider standard adiabatic
initial conditions, for which the ratio $\de\rho_i/\rho_i$ is of the same order
of magnitude for matter, radiation and k-essence. Since the Bardeen potential $\Psi$ is
related to $(\sum_i\de\rho_i)/\rho_{\rm tot}$, it is sourced mainly by the
perturbations in the dominant component of the universe. Therefore,
when k-essence is subdominant, we can write the equation of motion for
k-essence perturbations considering the Bardeen potential as an external source,
which does not influence the propagation properties. This equation is of the form
\be
\label{eq:pert_1}
\ddot{\delta\phi}+\al\dot{\de\phi}+\beta\de\phi+c_s^2\Delta\de\phi=\mu\dot{\Psi}+\nu\Psi~,
\ee
where $\Delta\de\phi=g^{ij}\dd_i\dd_j(\de\phi)$. Here the over-dot denotes
derivative with respect to physical time $t$ and $\al,\beta,c_s^2,\mu$ and $\nu$
are functions of $t$. Similar perturbation equations have also been
derived in~\cite{mac}. This equation is of the type (\ref{eqmo}). The
wave fronts 
are given by the characteristics, which determine the maximal speed of signal
propagation, here $c_s$. This sound velocity is achieved in the limit of high
wave-numbers $k\rightarrow \infty$, and  is given by
\be
c_s^2=\frac{p'}{2Xp''+p'}\hspace{1cm}'=\frac{d}{dX}~.
\ee
For the effective mass term $\beta$, we find
\be
\beta=\frac{2\rho_k}{2Xp''+p'}~,
\ee
which is always positive since  the energy density of k-essence,
$\rho_k=(2Xp'-p)/\phi^2$,  is positive,
and  $2Xp''+p'>0$ in a stable theory~\cite{ke+}. For the damping
term $\alpha\dot{\de\phi}$ we find
\be
\alpha=3H\left( 1-\frac{2Xp'(3p''+2Xp''')}{(p'+2Xp'')^2}\right)-2\frac{\dot{\phi}}{\phi}\left( 1+\frac{(p-2Xp')(3p''+2Xp''')}{(p'+2Xp'')^2}\right)~.
\ee
For illustrative purpose, we can now calculate the group velocity using a WKB approximation.
For simplicity we neglect the source term which does not affect the propagation
properties. We set
\be
\de\phi(x,t)=\int d^3k\, e^{ikx}\varphi(k,t)~,
\ee
where $t, k$ are the physical time and momentum. The Fourier
transformed function satisfies the equation
\be\label{pertdphi}
\ddot{\varphi}+\al\dot{\varphi}+(\beta+c_s^2k^2)\varphi=0~.
\ee
In order to put this equation in a form suitable for the WKB approximation, we
perform the substitution
\be
\varphi(k,t)=e^{-\int_0^t \frac{\al(t')}{2}dt'}A(k)W(t)~,
\ee
so that~(\ref{pertdphi}) reduces to
\be
\ddot{W}(t)+\om^2(k,t)W(t)=0~,
\ee
where we have identified
\be
\om^2(k,t)\equiv \beta+c_s^2k^2-\frac{\al^2}{4}-\frac{\dot{\al}}{2}~.
\ee
We define the effective mass term
\be
m^2\equiv \beta-\frac{\al^2}{4}-\frac{\dot{\al}}{2}~.
\ee

Within the WKB approximation we neglect the time derivatives of $c_s^2$
and of $m^2$, $\dot c_s/c_s\ll \om$ and $\dot m/m\ll \om $, yielding the
approximate solution
\be
\label{phifinal}
\de\phi(x,t)=e^{-\int_0^t \frac{\al}{2}dt'}\!
\int d^3k \frac{A(k)}{\sqrt{\om(k,t)}}e^{ikx-i\int_0^t\om(k,t')dt'}\, .
\ee
As customary in the evaluation of the group velocity, we now suppose that
$A(k)$ is a function sharply  peaked around a given wave-number $k_0$,
and that it stays so for at least a few oscillations.
We can therefore Taylor expand $\om(k,t)$ at first order both in $k-k_0$ and in $t$,
and within the WKB approximation we then find
\be
\de\phi(x,t)\simeq f(k_0,t)\,\de\phi(y,0)~,
\ee
where $f(k_0,t)$ is an irrelevant phase and
\be
y=x-\partial_k\om(k_0,0)t~.
\ee
The group velocity is therefore given by
\be
\label{vgroupWKB}
v_g(k_0)=\partial_k\om(k_0,0)=\frac{c_s^2k_0}{\sqrt{c_s^2k_0^2+m^2}}~.
\ee
If $m^2$ is positive, the velocity of the perturbation is always smaller
than $c_s$, and approaches it in the limit $k_0\rightarrow \infty$. If $m^2$ is
negative, low wave numbers  with $c_s^2k^2_0< -m^2$
are unstable. Because of the properties of hyperbolic equations of
motion~ \cite{CH}, we know that the maximal 
speed of the signal is again $c_s$. Therefore in this case, Eq.~(\ref{vgroupWKB})
no longer correctly describes the signal propagation speed.

\FIGURE[t]{\epsfig{file=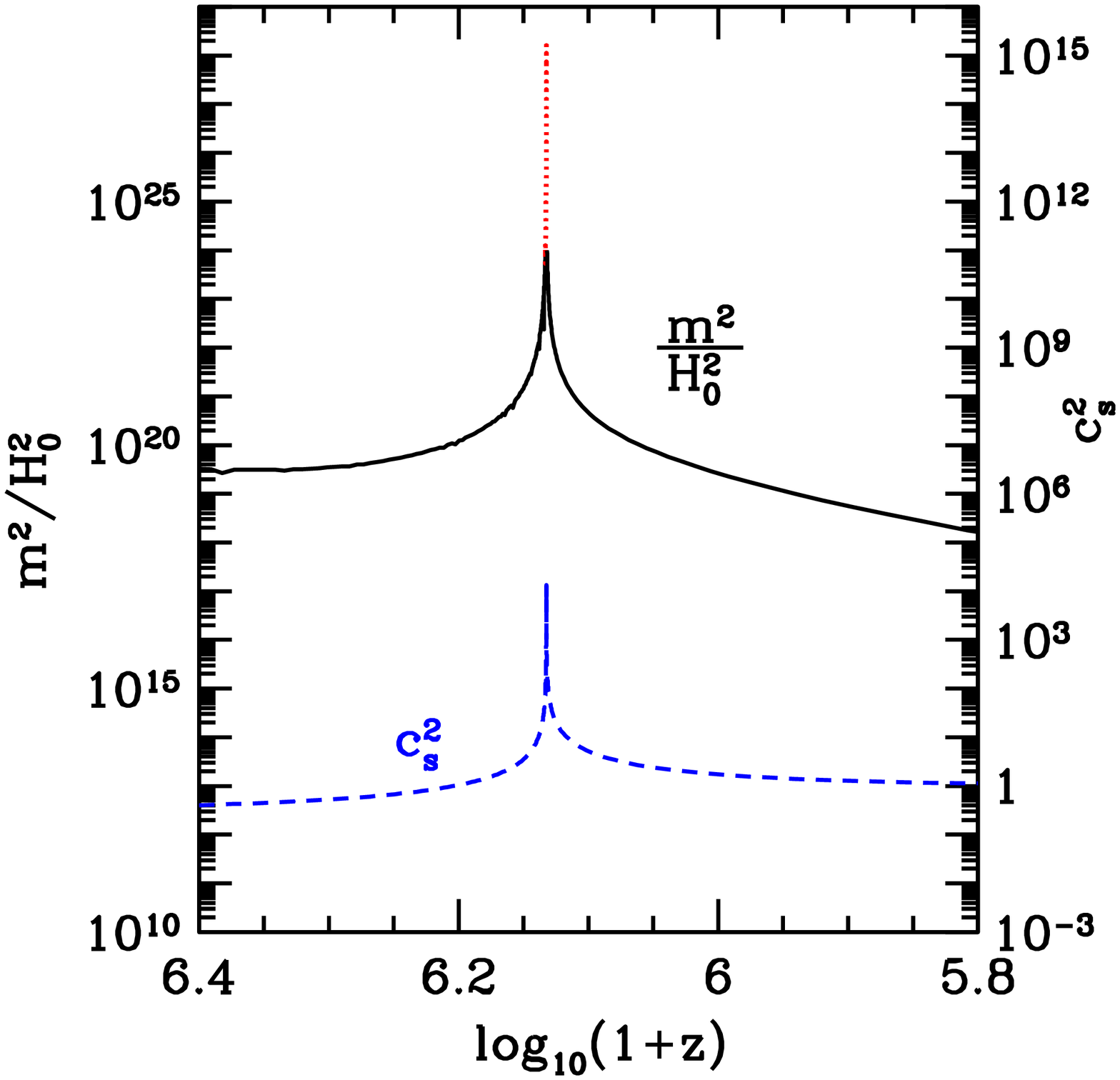,height=7.2cm}
\caption{\label{fig3}
The mass term $m^2$ and the sound velocity $c_s^2$ as functions of redshift for the example (\ref{example}).
We have plotted the absolute value of $m^2$ and the solid line corresponds to a positive $m^2$
whereas the dotted line corresponds to a negative $m^2$.}}

In Fig.~\ref{fig3}, we plot $m^2$ and $c_s^2$ for the k-essence Lagrangian given in
ref. \cite{ke1} 
\be 
\label{example}
p(X)=-2.01+2\sqrt{1+X}+3\cdot 10^{-17}X^3-10^{-24}X^4~.
\ee
We see that the condition $m^2>0$ (solid line) is verified for most of
the region of interest given by $c_s^2>1$ (dashed line). In the
example considered, $c_s^2>1$ after $z\simeq 1.4\times 10^6$ and stays so
until today~\cite{kess}. Note that the part where $m^2<0$ (dotted
line) corresponds to a stage where the background 
varies so quickly that in any case the WKB approximation
breaks down, and our calculation does not apply any longer.
If $m^2>0$, the group velocity is given by
equation (\ref{vgroupWKB}) for sufficiently large values of $k_0$ . In
order to construct the CSC of Fig.~\ref{fig1}, we now simply need to 
choose $k_1>k_2$ in order to have $v_1=v_g(k_1)>v_2=v_g(k_2)$, and $k_1,k_2$
large enough to have $v_1, v_2>1/v>1$.

The situation is analogous if we choose
non-adiabatic initial conditions where k-essence perturbations
are much larger than matter and radiation perturbations. Of course
the sound velocity $c_s^2$ which only depends on the second order spatial
derivatives in the equation of motion will remain the same. Therefore the
fact that the theory has a speed of sound larger than the speed of light
does not depend on the particular choice of the initial conditions. But
the group velocity can be different in the two cases.

Combining the three Einstein equations $00, 0i$ and $ii$, which relate
the evolution of the k-essence perturbations $\de\phi$ to the Bardeen
potential $\Psi$, we obtain a second order equation of motion for
$\Psi$ which has the form
\be
\label{eq:pert_2}
\ddot{\Psi}+\tilde{\al}\dot{\Psi}+\tilde{\beta}\Psi+c_s^2\Delta\Psi=0~,
\ee
where now 
\be
\tilde{\beta}=2\dot{H}+2\frac{H}{p'+2Xp''}\Bigg(3H(p'+Xp'')-\frac{2g(X)}{p'\phi\dot{\phi}} \Bigg)\,
\ee
and the damping term
\be
\tilde{\al}=\frac{7p'+8Xp''}{p'+2Xp''}H-\frac{4g(X)}{(p'+2Xp'')p'\phi\dot{\phi}}~,
\ee
with $g(X)=Xp'^2-pp'-Xpp''$.

The above calculation of the group velocity can be straightforwardly repeated
in this case. One finds for the Bardeen potential the same form of the
group velocity as in (\ref{vgroupWKB}), but in terms of the new effective mass $\tilde m$
given by
$\tilde{m}^2\equiv \tilde{\beta}-\frac{\tilde{\al}^2}{4}-\frac{\dot{\tilde{\al}}}{2}$.
As in the previous case, we have evaluated $\tilde{m}^2$ for the particular Lagrangian (\ref{example}), 
and we find the same qualitative behaviour as for $m^2$ in Fig. \ref{fig3}.  

\section{Conclusions}

We have shown that if superluminal motion is possible and if a
signal emitted in some reference frame
$R'$  propagates always forward with respect to the frame time
$t'$, closed signal curves, CSC's can be constructed. Note that these
are neither closed timelike curves nor closed causal curves (timelike
or lightlike) in the
sense of Hawking and Ellis~\cite{HE}, since they contain spacelike
parts. Hawking and Ellis call spacetimes which do not admit closed
causal curves 'causally stable', and they show that stable
causality is equivalent to the existence of a Lorentzian metric and of a
function $t$ the gradient of which is globally timelike and past-directed~\cite{HE},
p198ff. This condition may very well be satisfied in our case since
the field $\phi$ may be weak and the metric nearly flat.

However the relevant question is whether the existence of a global
past-directed timelike gradient $\nabla^\alpha t$ prevents also the
existence of closed signal curves which are partially spacelike, 
as constructed in Fig.~\ref{fig1}. 
As argued in ~\cite{HE} (see also~\cite{Wald}), if a past-directed
timelike gradient $\nabla^\alpha t$ exists, closed timelike or lightlike curves
cannot be formed since for every future-directed timelike or
lightlike curve with tangent $v^\alpha$ the derivative of $t$ along
the curve $g_{\alpha\beta}v^\alpha\nabla^\beta t <0$.  
This means that $t$ can only decrease along such a curve and therefore
can never return to its initial value.

The situation is different if one allows a signal to propagate along a
spacelike curve, even if it remains inside a given cone defined by a
Lorentzian metric $G_{\alpha\beta}$. 
Indeed the notion of 'future-directed curve' is not well defined for a
spacelike curve; it depends on the reference frame. Therefore we
cannot apply the same argument as before; first we have to choose a
notion of 'future-directed' for spacelike curves. Let us first use the
notion which has led to the CSC:
We define (unambiguously) a curve to be  future-directed, if 
a signal along this curve always propagates forward in time with
respect to the reference frame in which it has been emitted. With this
definition the curve from $q_0$ to $q_1$ as well as the one from $q_1$
to $q_2$ are both future-directed. But, denoting by $v_1^\alpha$ and
$v_2^\alpha$ their tangent vectors, we clearly have
$g_{\alpha\beta}v_1^\alpha\nabla^\beta t<0$ but
$g_{\alpha\beta}v_2^\alpha\nabla^\beta t>0$ and
$g_{\alpha\beta}v_1^\alpha\nabla^\beta t' >0$ but
$g_{\alpha\beta}v_2^\alpha\nabla^\beta t'<0$. 
Every timelike coordinate which grows along one part of our curve,
decays along the other part. Therefore the conditions of the 
theorem are no longer met and one can construct closed signal curves.

On the other hand if we introduce a preferred frame and require
that signals always propagate forward in time with respect to this frame,
we can define the notion of future-directed curve in this frame and
the theorem does apply: no closed curves can be constructed. But the
price to pay is that in other reference frames emitters can send
signal which are past-directed with respect to their proper time.
In the reference frame with velocity $v=1/v_1$ the signal even
propagates with infinite velocity, which means that the 'propagation
equation' is no longer hyperbolic but elliptic. In this frame the
propagation of the fluctuations $\de\phi$ becomes non-local. 

Finally, one may consider to apply the  Hawking and Ellis theorem not 
to the spacetime metric $g_{\alpha\beta}$ but to the metric
$G_{\alpha\beta}$. In 
this case even if a curve has a superluminal velocity, it can be a timelike 
future-directed curve with respect to $G_{\alpha\beta}$ and therefore
$G_{\alpha\beta}v^\alpha\nabla^\beta t< 0$, which implies that no
closed timelike (with respect to $G_{\alpha\beta}$) curve
can be formed. But this notion is invariant only with respect to 'Lorentz
transformations' which leave $G_{\alpha\beta}$ invariant
and not the lightcone. Therefore, now the speed of light depends on
the reference frame. Furthemore, local Lorentz symmetry with respect
to $G_{\alpha\beta}$ would now imply that we have to take covariant
derivatives with respect to this metric. 
Hence it is $G_{\alpha\beta}$ and no longer $g_{\alpha\beta}$ which
defines the structure of spacetime, and we replace general relativity
by a bi-metric theory of gravity.
 
Hence the Hawking and Ellis theorem confirms our conclusions: if
superluminal motion is possible and if a signal emitted in some
reference frame $R'$  propagates always forward in frame-time
$t'$, closed signal curves can be constructed. These curves, even if 
they are not timelike or lightlike, are 'time machines'. They allow us,
e.g., to influence the present with knowledge of the future. After
watching the 6 numbers on TV on Saturday evening we can send this
information back to Friday afternoon and enter them in our lottery
bulletin.

On the other hand, if a background which defines a preferred timelike
direction is present, the ruin of all lottery companies can sometimes be
prevented: we just have to require that signals travel
forward in time in a preferred rest frame which can be
defined unambiguously if $\nabla\phi_0$ is timelike.
But this implies that in other reference frames
a signal can propagate either towards the future or towards the past,
depending on the direction of emission (or it can even behave
non-locally). 

As we have started with a Lorentz invariant
Lagrangian, we would in principle expect that all solutions of the
equations of motion are viable and that their perturbations 
have to be handled in the same way. However in order to avoid CSC's,
theories that allow for superluminal motion have, in addition to the
Lagrangian, to provide a rule which tells us when to take the 
retarded and when the advanced Green function to propagate perturbations
on a background. If the background solution has no special symmetries,
it is not straightforward to implement such a rule.

\acknowledgments{It is a pleasure to thank Martin Kunz, Norbert
 Straumann, Riccardo  Sturani and Alex Vikman for stimulating
 discussions. We thank Marc-Olivier Bettler for his help with the
 figures.  This work is supported by the
 Swiss National Science Foundation.}

\end{document}